\documentclass[pre,twocolumn,showpacs,amsmath,amssymb]{revtex4}

\usepackage{natbib}
\usepackage{epsfig}
\usepackage{amsmath,amssymb,mathrsfs}

\newcommand{\calA}{\mathcal A}
\newcommand{\calC}{\mathcal C}
\newcommand{\calP}{\mathcal P}
\newcommand{\Laplace}{\hat{\mathscr L}}
\newcommand{\rrr}{\mathbf{r}}
\newcommand{\rrrp}{\mathbf{r}_\perp}
\newcommand{\nablabf}{\boldsymbol{\nabla}}
\newcommand{\ex}{\mathbf{e^{{}}}_x}
\newcommand{\ey}{\mathbf{e^{{}}}_y}
\newcommand{\ez}{\mathbf{e^{{}}}_z}
\newcommand{\vvv}{\mathbf{v^{{}}}}

\begin{document}

\title{Universal dynamics in the onset of a Hagen--Poiseuille flow}

\author{Niels Asger Mortensen and Henrik Bruus}

\affiliation{MIC -- Department of Micro and Nanotechnology, NanoDTU,\\
Technical University of Denmark, DK-2800 Kongens Lyngby, Denmark }

\date{January 17, 2006}

\begin{abstract}
The dynamics in the onset of a Hagen--Poiseuille flow of an
incompressible liquid in a channel of circular cross section is
well-studied theoretically. We use an eigenfunction expansion in a
Hilbert space formalism to generalize the results to channels of
arbitrary cross section. We find that the steady state is reached
after a characteristic time scale $\tau = (\calA/\calP)^2 (1/\nu)$
where $\calA$ and $\calP$ are the cross-sectional area and
perimeter, respectively, and $\nu$ is the kinematic viscosity of
the liquid. For the initial dynamics of the flow rate $Q$ for
$t\ll\tau$ we find a universal linear dependence, $Q(t)=
Q_\infty\: (\alpha/\calC)\: (t/\tau)$, where $Q_\infty$ is the
asymptotic steady-state flow rate, $\alpha$ is the geometrical
correction factor, and $\calC=\calP^2/\calA$ is the compactness
parameter. For the long-time dynamics $Q(t)$ approaches $Q_\infty$
exponentially on the timescale $\tau$, but with a weakly
geometry-dependent prefactor of order unity, determined by the
lowest eigenvalue of the Helmholz equation.
\end{abstract}


 \maketitle

\section{Introduction}

Hagen--Poiseuille flow (or simply Poiseuille flow) is important to
a variety of applications ranging from macroscopic pipes in
chemical plants to the flow of blood in veins. However, the rapid
development in the field of lab-on-a-chip systems during the past
decade has put even more emphasis on pressure driven laminar flow.
Traditionally, capillary tubes would have circular cross-sections,
but today microfabricated channels come with a variety of shapes
depending on the fabrication technique in use. The list of
examples includes rectangular channels obtained by hot embossing
in polymer wafers, semi-circular channels in isotropically etched
surfaces, triangular channels in KOH-etched silicon crystals,
Gaussian-shaped channels in laser-ablated polymer films, and
elliptic channels in stretched PDMS devices, see e.g.
Ref.~\onlinecite{Geschke:04a}.

This development has naturally led to more emphasis on theoretical
studies of shape-dependence in microfluidic channels. Recently,
 we therefore revisited the problem of Poiseuille flow and
its shape dependence and we have also addressed mass diffusion in
microchannels~\cite{Mortensen:05b,Mortensen:05c}. In the present
work we combine the two former studies and address the dynamics
caused by the abrupt onset of a pressure gradient at time $t=0$ in
an incompressible liquid of viscosity $\eta$ and density $\rho$
situated in a long, straight, and rigid channel of length $L$ and
some constant cross-sectional shape $\Omega$. The solution is
well-known for the case of a cylindrical channel, see e.g.
Ref.~\onlinecite{Batchelor:67}, but in this paper we generalize
the results to a cross-section of arbitrary shape. The similarity
between mass and momentum diffusion, and the existence of a
characteristic diffusion time-scale $\tau^{{}}_\textrm{diff} =
(\pi/4)(\calA/\calP)^2/D$ for mass diffusion~\cite{Mortensen:05c}
have led us to introduce the momentum diffusion time-scale $\tau$
defined by
 \begin{equation} \label{eq:tau}
 \tau=\Big(\frac{\calA}{\calP}\Big)^2\frac{1}{\nu},
 \end{equation}
where $\nu=\eta/\rho$ is the kinematic viscosity (having
dimensions of a diffusion constant), while $\calA$ and $\calP$ is
the area and perimeter of the cross section $\Omega$,
respectively. In this paper we show that the dynamics of the flow
rate $Q(t)$ is universal with $\tau$ as the characteristic time
scale.

As shown in Ref.~\onlinecite{Mortensen:05b} the shape parameters
$\calA$ and $\calP$ also play an important role in the
steady-state Poiseuille flow. The hydraulic resistance $R_{\rm
hyd}$ can be expressed as
 \begin{equation} \label{eq:Rhyd}
 R_{\rm hyd} = \alpha\:\frac{\eta L}{\calA^2} \equiv
 \alpha R_{\rm hyd}^*,
 \end{equation}
where $\alpha$ is a dimensionless geometrical correction factor
and $R_{\rm hyd}^*=\eta L/\calA^2$ is a characteristic resistance.
Remarkably, $\alpha$ is simply (linearly) related to the
dimensionless compactness parameter $\calC=\calP^2/\calA$.

Above we have emphasized microfluidic flows because of the variety
of shapes frequently encountered in lab-on-a-chip systems.
However, our results are generally valid for laminar flows at any
length scale.

\section{Diffusion of momentum}

We consider a long, straight channel of length $L$, aligned with
the $z$-axis, having a constant cross section $\Omega$ with the
boundary $\partial\Omega$ in the $xy$ plane. The fluid flow is
driven by a pressure gradient of $\nablabf p = -(\Delta p/L)\ex$
which is turned on abruptly at time $t=0$. We note that strictly
speaking the pressure gradient is not established instantaneously,
but rather on a time-scale set by $L/c$ where $c$ is the speed of
sound. For typical liquids $c\sim 10^3\,{\rm m/s}$ which for
micro-fluidic systems and practical purposes makes the pressure
gradient appear almost instantaneously. From the symmetry of the
problem it follows that the velocity field is of the form
$\vvv(\rrr,t)=v(\rrrp,t)\ex$ where $\rrrp = y\ey+z\ez$. From the
Navier--Stokes equation it then follows that $v(\rrrp,t)$ is
governed by, see e.g. Refs.~\onlinecite{Batchelor:67}
or~\onlinecite{Landau:87a},
 \begin{equation}\label{eq:momentumdiffusion}
 \frac{1}{\nu} \:\partial_t v(\rrrp,t)- \nabla^2
 v(\rrrp,t)=\frac{\Delta p}{\eta L},
 \end{equation}
which is a diffusion equation for the momentum with the pressure
drop acting as a source term on the right-hand side. The velocity
$v$ is subject to a no-slip boundary condition on $\partial\Omega$
and obviously $v$ is initially zero, while it asymptotically
approaches the steady-state velocity field $v_\infty(\rrrp)$ for
$t\rightarrow\infty$.

In the analysis it is natural to write the velocity as a
difference
 \begin{equation}
 \label{eq:vhvinfty}
 v(\rrrp,t) = v_\infty(\rrrp) - v^{{}}_h(\rrrp,t)
 \end{equation}
of the asymptotic, static field $v_\infty$, solving the Poiseuille
problem
 \begin{equation}\label{eq:Poiseuille}
 - \nabla^2 v_\infty(\rrrp)=\frac{\Delta p}{\eta L},
 \end{equation}
and a time-dependent field $v^{{}}_h(\rrrp,t)$ satisfying the
homogeneous diffusion equation,
 \begin{equation} \label{eq:vhHelmholz}
 \frac{1}{\nu} \:\partial_t v^{{}}_h(\rrrp,t)- \nabla^2 v^{{}}_h(\rrrp,t)= 0.
 \end{equation}
From Ref.~\onlinecite{Mortensen:05c} it is known that rescaling
the Helmholz equation by $(\calA/\calP)^2$ leads to a lowest
eigenvalue $a_1$ that is of order unity and only weakly geometry
dependent. We therefore perform this rescaling, which naturally
implies the time-scale $\tau$ of Eq.~(\ref{eq:tau}) and the
following form of the diffusion equation,
 \begin{equation}
 \label{eq:HelmholzRescaled}
 \tau \:\partial_t v^{{}}_h(\rrrp,t) -
\Laplace v^{{}}_h(\rrrp,t)= 0.
 \end{equation}
where we have introduced the rescaled Laplacian $\Laplace$,
 \begin{equation}
 \label{eq:LaplaceRescaled}
 \Laplace \equiv \Big(\frac{\calA}{\calP}\Big)^2 \nabla^2.
 \end{equation}
We note that by the rescaling the Navier--Stokes
equation~(\ref{eq:momentumdiffusion}) becomes
 \begin{equation}
 \label{eq:NSrescaled}
 \tau \:\partial_t v - \Laplace v
 = \Big(\frac{\calA}{\calP}\Big)^2 \frac{\Delta p}{\eta L}
 = \frac{\alpha Q_\infty}{\calP^2},
 \end{equation}
where we have introduced the steady-state flow rate $Q_\infty =
\Delta p/R_\textrm{hyd}$ and used Eq.~(\ref{eq:Rhyd}).

\section{Hilbert space formulation}
In order to solve Eq.~(\ref{eq:NSrescaled}) we will take advantage
of the Hilbert space formulation~\cite{Morse:1953}, often employed
in quantum mechanics~\cite{Merzbacher:70}. The Hilbert space of
real functions $f(\rrrp)$ is defined by the inner product
 \begin{equation}
 \big< f \big|g\big>\equiv \int_\Omega d\rrrp\, f(\rrrp)g(\rrrp)
 \end{equation}
and a complete set $\big\{\big|\phi_n\big>\big\}$ of orthonormal
basis functions,
 \begin{equation}
 \big<\phi_m\big|\phi_n\big>=\delta_{nm}.
 \end{equation}
Above, we have used the Dirac \emph{bra-ket} notation and
$\delta_{nm}$ is the Kronecker delta. We choose the eigenfunctions
$\{\big|\phi_n\big>\}$ of the rescaled Helmholz equation (with a
zero Dirichlet boundary condition on $\partial\Omega$) as our
basis functions,
 \begin{equation}
 \label{eq:LaplaceInv}
 -\Laplace\big|\phi_n\big>=a_n\big|\phi_n\big>.
 \end{equation}
With this complete basis any function in the Hilbert space can be
written as a linear combination of basis functions. Using the
\emph{bra-ket} notation Eq.~(\ref{eq:NSrescaled}) becomes
 \begin{equation}\label{eq:momentumdiffusion_braket}
 \tau \:\partial_t \big|v\big>- \Laplace
 \big|v\big>=\frac{\alpha Q_\infty}{\calP^2}\big|1\big>.
 \end{equation}
The full solution Eq.~(\ref{eq:vhvinfty}) is written as
 \begin{equation}\label{eq:Ansatz}
 \big|v\big>=\big|v_\infty\big>-\big|v^{{}}_h\big>,
 \end{equation}
where $\big|v_\infty\big>$ satisfies the Poiseuille problem
Eq.~(\ref{eq:Poiseuille}),
 \begin{equation}\label{eq:VinftyDef}
 -\Laplace\big|v_\infty\big>=\frac{\alpha Q_\infty}{\calP^2}\big|1\big>,
 \end{equation}
and the homogeneous solution $\big|v^{{}}_h\big>$ solves the
diffusion problem Eq.~(\ref{eq:HelmholzRescaled})
 \begin{equation}
 \big(\tau\partial_t-\Laplace\big)\big|v^{{}}_h\big>=0.
 \end{equation}
In the complete basis $\{\big|\phi_n\big>\}$ we have
 \begin{align}
 \big|v^{{}}_h\big>&=\sum_{n=1}^{\infty} b_n e^{-a_n t/\tau}\big|\phi_n\big>,\label{eq:vh}\\
 \big|v_\infty\big>&=\sum_{n=1}^{\infty} c_n \big|\phi_n\big>,\label{eq:vinfty}
 \end{align}
and since $\lim_{t\rightarrow
0}\big|v^{{}}_h\big>=\big|v_\infty\big>$ we have $b_n=c_n$.
Multiplying Eq.~(\ref{eq:vinfty}) by $\big<\phi_m\big|$ yields
 \begin{align}
 b_m = c_m &= \big<\phi_m\big|v_\infty\big>=
 \big<\phi_m\big|\Laplace^{-1}\Laplace\big|v_\infty\big>\nonumber\\
 &=
 \frac{\alpha Q_\infty}{\calP^2}\:a_m^{-1} \big<\phi_m\big|1\big>.\label{eq:bncn}
 \end{align}
In the second-last equality we have introduced the unit operator
$1=\Laplace^{-1}\Laplace$ and in the last equality we used the
Hermitian property of the inverse Laplacian operator to let
$\Laplace^{-1}$ act to the left, $\big<\phi_m\big|\Laplace^{-1} =
-\big<\phi_m\big|a_m^{-1}$ from Eq.~(\ref{eq:LaplaceInv}), while
$\Laplace$ acts to the right, see Eq.~(\ref{eq:VinftyDef}).
Substituting Eqs.~(\ref{eq:vh}) and~(\ref{eq:bncn}) into
Eq.~(\ref{eq:Ansatz}) we finally obtain
 \begin{equation} \label{eq:Vexpansion}
 \big|v\big> = \big|v_\infty\big> -
 \frac{\alpha Q_\infty}{\calP^2}\sum_{n=1}^{\infty} \big|\phi_n\big>
 \big<\phi_n\big|1\big>\:a_n^{-1} e^{-a_n t/\tau}.
 \end{equation}

\section{Flow rate}
Using the \emph{bra-ket} notation, the flow rate $Q(t)$ at any
time is conveniently written as $Q=\big<1\big|v\big>$, and thus in
steady state $Q_\infty=\big<1\big|v_\infty\big>$. Multiplying
Eq.~(\ref{eq:Vexpansion}) from the left by $\big<1\big|$ yields
 \begin{equation}\label{eq:Qbraket}
 Q(t) = \big<1\big|v\big> =
 Q_\infty - \frac{\alpha Q_\infty}{\calP^2}
 \sum_{n=1}^{\infty} \big< 1 \big|\phi_n\big>
 \big<\phi_n\big|1\big>\:a_n^{-1} e^{-a_n t/\tau}.
 \end{equation}
The factor $\big< 1 \big|\phi_n\big> \big<\phi_n\big|1\big>$ is
recognized as the effective area $\calA_n$ covered by the $n$th
eigenfunction $\big|\phi_n\big>$,
 \begin{equation}
 \calA_n\equiv
 \frac{\big|\big<1\big|\phi_n\big>\big|^2}{
 \big<\phi_n\big|\phi_n\big>}
 = \big|\big<1\big|\phi_n\big>\big|^2
 = \big< 1 \big|\phi_n\big> \big<\phi_n\big|1\big>.
 \end{equation}
The effective areas fulfil the sum-rule $\sum_{n=1}^{\infty}
\calA_n = \calA$, seen by completeness of the basis
$\big\{\big|\phi_n\big>\big\}$.
%
%
%
We can find the geometrical correction factor $\alpha$ from
Eq.~(\ref{eq:Qbraket}) by using that $Q(0) = 0$,
 \begin{equation}\label{eq:alpha}
 \alpha =\calP^2
 \Bigg(\sum_{n=1}^{\infty} \frac{\calA_n}{a_n}\Bigg)^{-1},
 \end{equation}
and substituting into Eq.~(\ref{eq:Qbraket}) we finally get
 \begin{equation}\label{eq:Q_general}
 \frac{Q(t)}{Q_\infty}=1-
 \Bigg(\sum_{n=1}^{\infty} \frac{\calA_n}{a_n}\Bigg)^{-1}
 \sum_{n=1}^{\infty} \frac{\calA_n}{a_n}\: e^{-a_n t/\tau}.
\end{equation}

\section{Short-time dynamics}

The short-time dynamics is found by Taylor-expanding
Eq.~(\ref{eq:Q_general}) to first order,
 \begin{equation}
 \frac{Q(t)}{Q_\infty} \approx
 \Bigg(\sum_{n=1}^{\infty} \frac{\calA_n}{a_n}\Bigg)^{-1}
 \calA\: \frac{t}{\tau}
 = \frac{\alpha\calA}{\calP^2}\: \frac{t}{\tau}
 = \frac{\alpha}{\calC}\:\frac{t}{\tau},
 \quad t\ll\tau,
 \end{equation}
where we have used the sum-rule for $\calA_n$ as well as
Eq.~(\ref{eq:alpha}). The short time dynamics can also be inferred
directly by integration of the Navier--Stokes equation
Eq.~(\ref{eq:momentumdiffusion_braket}), since at time $t=0$ we
have $\big|v\big>=0$ and consequently the vanishing of velocity
gradients and viscous friction, $\Laplace \big|v\big>=0$. Thus we
arrive at
 \begin{equation}
 \tau \:\partial_t \big|v\big> =
 \frac{\alpha Q_\infty}{\calP^2}\big|1\big>,
 \quad t\rightarrow 0,
 \end{equation}
corresponding to a constant initial acceleration throughout the
fluid. Integration with respect to $t$ is straightforward and
multiplying the resulting $\big|v\big>$ by $\big<1\big|$ yields
$Q(t)$,
 \begin{equation}\label{eq:Qshort}
 \frac{Q(t)}{Q_\infty}
 \simeq \frac{\alpha}{\calP^2}\big<1\big|1\big> \frac{t}{\tau}
 = \frac{\alpha \calA}{\calP^2} \frac{t}{\tau}
 = \frac{\alpha}{\calC} \: \frac{t}{\tau},\quad
 t\ll\tau.
\end{equation}
Thus initially, the fluid responds to the pressure gradient in the
same way as a rigid body responds to a constant force.

\begin{table}[t!]
\begin{center}
 \begin{tabular}{lccccc}
 shape & $a_1$ & $\calA_1/\calA$&$\alpha/\calC$\\\\\hline
 circle & $\gamma_{0,1}^2/4\simeq 1.45$$^a$ & $4/\gamma_{0,1}^2\simeq 0.69$$^a$ & 2$^b$\\
 quarter-circle  & 1.27$^c$ & 0.65$^c$ & 1.85$^c$ \\
 half-circle     & 1.38$^c$ & 0.64$^c$ & 1.97$^c$\\
 ellipse(1:2)    & 1.50$^c$ & 0.67$^c$ & 2.10$^c$\\
 ellipse(1:3)    & 1.54$^c$ & 0.62$^c$ & 2.21$^c$\\
 ellipse(1:4)    & 1.57$^c$ & 0.58$^c$ & 2.28$^c$\\\hline
 triangle(1:1:1) & $\pi^2/9\simeq 1.10$$^d$& $6/\pi^2\simeq 0.61$$^d$ & $5/3$$^b$\\
 triangle(1:1:$\sqrt{2}$) & $\frac{5\pi^2}
  {4( 2 + \sqrt{2})^2}\simeq 1.06$$^a$ &$512/9\pi^4\simeq 0.58$$^a$
  &1.64$^c$\\\hline
 square & $\pi^2/8\simeq 1.23$$^a$ & $64/\pi^4\simeq 0.66$$^a$ & 1.78$^c$\\
 rectangle(1:2) & $5\pi^2/36\simeq 1.37$$^a$& $64/\pi^4\simeq 0.66$$^a$ & 1.94$^c$\\
 rectangle(1:3) & $ 5\pi^2/32\simeq 1.54$$^a$ & $64/\pi^4\simeq 0.66$$^a$ & 2.14$^c$\\
 rectangle(1:4) & $17\pi^2/100\simeq 1.68$$^a$ &$64/\pi^4\simeq 0.66$$^a$ & 2.28$^c$ \\
 rectangle(1:$\infty$) & $ \sim\pi^2/4\simeq 2.47$$^a$
 &$64/\pi^4\simeq 0.66$$^a$ & $\sim 3$$^e$\\\hline
 pentagon &  1.30$^c$ & 0.67$^c$ &1.84$^c$\\\hline
 hexagon & 1.34$^c$ & 0.68$^c$ & 1.88$^c$\\\hline
 \end{tabular}
\caption{Central parameters for the lowest eigenfunction for
different cross sectional shapes. Note how the different numbers
converge when going through the regular polygons starting from the
triangle(1:1:1) through the square, the pentagon, and the hexagon
to the circle. $^a$See e.g. Ref.~\onlinecite{Morse:1953} for the
eigenmodes and eigenspectrum. Here, $\gamma_{0,1}$ denotes the
first root of the zeroth Bessel function of the first kind.
$^b$See Ref.~\onlinecite{Mortensen:05b}. $^c$Data obtained by
finite-element simulations. $^d$See e.g.
Ref.~\onlinecite{Brack:1997} for the eigenmodes and eigenspectrum.
$^e$See e.g. Ref.~\onlinecite{Batchelor:67} for a solution of the
Poisson equation.} \label{tab:1}
\end{center}
\end{table}

\section{Long-time dynamics}

As the flow rate increases, friction sets in, and in the long-time
limit $t\gg\tau$ the flow-rate saturates at the value $Q_\infty$
where there is a balance between the pressure gradient and
frictional forces. For the long-time saturation dynamics the
lowest eigenstate plays the dominating role and taking only the
$n=1$ term in Eq.~(\ref{eq:Q_general}) we obtain
\begin{equation}
\frac{Q(t)}{Q_\infty}  \simeq 1- e^{-a_1 t/\tau},\quad t\gg
\tau/a_2,\label{eq:Qlong}
\end{equation}
where we have used that the lowest eigenvalue $a_1$ is
non-degenerate to truncate the summation.

The time it takes to reach steady-state is denoted $\tau_\infty$.
A lower bound $\tau_1$ for $\tau_\infty$ can be obtained from
Eq.~(\ref{eq:Qshort}) by assuming that the initial acceleration is
maintained until $Q(\tau_1)/Q_\infty = 1$ is reached,
 \begin{equation} \label{eq:tau1}
 \tau_\infty > \tau_1 = \frac{\calC}{\alpha}\:\tau.
 \end{equation}
A better estimate $\tau_2$ for $\tau_\infty$ is obtained from
Eq.~(\ref{eq:Qlong}) by demanding $Q(\tau_2)/Q_\infty = 1 -
e^{-3}$.
 \begin{equation} \label{eq:tau2}
 \tau_\infty \approx \tau_2 = \frac{3}{a_1}\:\tau.
 \end{equation}
Using the parameter values for the circle listed in
Table~\ref{tab:1} we find the values
 \begin{equation} \label{eq:tauvalues}
 \tau_1 = 0.5\:\tau < \tau_2 = 2.1\:\tau \approx \tau_\infty.
 \end{equation}

\begin{figure}
\centerline{\epsfig{file=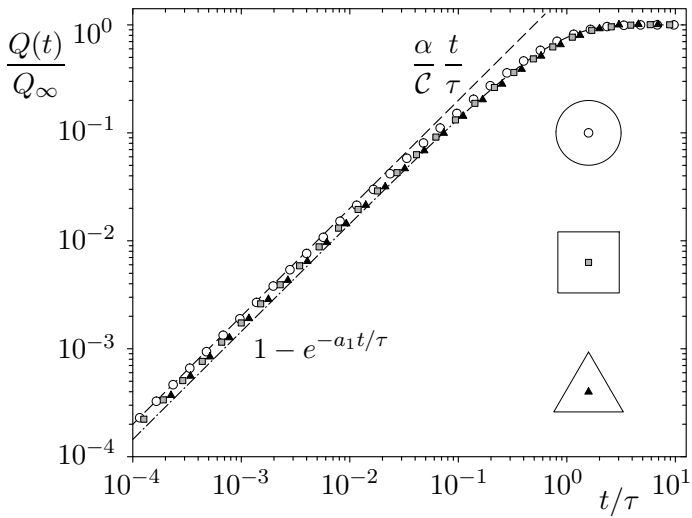}} \caption{A log-log
plot of the flow rate $Q(t)/Q_\infty$ as a function of time
$t/\tau$. The dashed line is the short-time approximation
Eq.~(\ref{eq:Qshort}), while the dashed-dotted line is the
long-time approximation Eq.~(\ref{eq:Qlong}), both for the case of
a circle, i.e., using $\calC/\alpha = 2$ and $a_1 = 1.45$ as
listed in Table~\ref{tab:1}. The data points are the results of
time-dependent finite-element simulations for the cases of the
cross section being a circle (white circles), a square (gray
squares), and an equilateral triangle (black triangles).
\label{fig1} }
\end{figure}

\section{Numerical results}

Only few geometries allow analytical solutions of both the
Helmholz equation and the Poisson equation. The circle is of
course the most well-known example, but the equilateral triangle
is another exception. However, in general the equations have to be
solved numerically, and for this purpose we have used the
commercially available finite-element software Comsol~3.2 (see
www.comsol.com). Numbers for a selection of geometries are
tabulated in Table~\ref{tab:1}.

The circle is the most compact shape and consequently it has the
largest value for $\calA_1/\calA$, i.e., the mode has the
relatively largest spatial occupation of the total area. The
eigenvalue $a_1$ is of the order unity for compact shapes and in
general it tends to increase slightly with increasing values of
$\calC$. The modest variation from geometry to geometry in both
$a_1$ and the other parameters suggests that the dynamics of
$Q(t)$ will appear almost universal.

In order to illustrate the validity of our two asymptotic
expressions, Eqs.~(\ref{eq:Qshort}) and~(\ref{eq:Qlong}), we have
compared them using the values for a circular shape to
time-dependent finite-element simulations of
Eq.~(\ref{eq:momentumdiffusion}). As illustrated in
Fig.~\ref{fig1} we find a perfect agreement between the asymptotic
expressions Eqs.~(\ref{eq:Qshort}) and~(\ref{eq:Qlong}) and the
numerically exact data for a circle, a square, and an equilateral
triangle. Comparing the corresponding parameters in
Table~\ref{tab:1} we would expect all data to almost coincide,
which is indeed also observed in Fig.~\ref{fig1}. The small spread
in eigenvalues and other parameters thus gives rise to
close-to-universal dynamics. From the plot it is also clear that
$\tau$ is indeed a good estimate for the time it takes to reach
the steady state.

\section{Conclusions}\label{sec:concl}

In conclusion, by using a compact Hilbert space formalism we have
shown how the initial dynamics in the onset of Poiseuille flow is
governed by a universal linear raise in flow rate $Q(t)$ over a
universal time-scale $\tau$ above which it saturates exponentially
to the steady-state value $Q_\infty$. The steady state is reached
after a time $\tau_\infty\approx{\calC}\tau/\alpha$. Apart from
being a fascinating example of universal dynamics for a complex
problem our results may have important applications in design of
real-time programmable pressure-driven micro-fluidic networks.


\end{document}